



\font\twelverm=cmr10 scaled 1200    \font\twelvei=cmmi10 scaled 1200
\font\twelvesy=cmsy10 scaled 1200   \font\twelveex=cmex10 scaled 1200
\font\twelvebf=cmbx10 scaled 1200   \font\twelvesl=cmsl10 scaled 1200
\font\twelvett=cmtt10 scaled 1200   \font\twelveit=cmti10 scaled 1200

\skewchar\twelvei='177   \skewchar\twelvesy='60


\def\twelvepoint{\normalbaselineskip=12.4pt
  \abovedisplayskip 12.4pt plus 3pt minus 9pt
  \belowdisplayskip 12.4pt plus 3pt minus 9pt
  \abovedisplayshortskip 0pt plus 3pt
  \belowdisplayshortskip 7.2pt plus 3pt minus 4pt
  \smallskipamount=3.6pt plus1.2pt minus1.2pt
  \medskipamount=7.2pt plus2.4pt minus2.4pt
  \bigskipamount=14.4pt plus4.8pt minus4.8pt
  \def\rm{\fam0\twelverm}          \def\it{\fam\itfam\twelveit}%
  \def\sl{\fam\slfam\twelvesl}     \def\bf{\fam\bffam\twelvebf}%
  \def\mit{\fam 1}                 \def\cal{\fam 2}%
  \def\tt{\twelvett}
  \def\nullspace{\nulldelimiterspace=0pt \mathsurround=0pt }
  \def\big##1{{\hbox{$\left##1\vbox to 10.2pt{}\right.\nullspace$}}}
  \def\Big##1{{\hbox{$\left##1\vbox to 13.8pt{}\right.\nullspace$}}}
  \def\bigg##1{{\hbox{$\left##1\vbox to 17.4pt{}\right.\nullspace$}}}
  \def\Bigg##1{{\hbox{$\left##1\vbox to 21.0pt{}\right.\nullspace$}}}
  \textfont0=\twelverm   \scriptfont0=\tenrm   \scriptscriptfont0=\sevenrm
  \textfont1=\twelvei    \scriptfont1=\teni    \scriptscriptfont1=\seveni
  \textfont2=\twelvesy   \scriptfont2=\tensy   \scriptscriptfont2=\sevensy
  \textfont3=\twelveex   \scriptfont3=\twelveex  \scriptscriptfont3=\twelveex
  \textfont\itfam=\twelveit
  \textfont\slfam=\twelvesl
  \textfont\bffam=\twelvebf \scriptfont\bffam=\tenbf
  \scriptscriptfont\bffam=\sevenbf
  \normalbaselines\rm}



\def\beginlinemode{\endmode
  \begingroup\parskip=0pt \obeylines\def\\{\par}\def\endmode{\par\endgroup}}
\def\beginparmode{\endmode
  \begingroup \def\endmode{\par\endgroup}}
\let\endmode=\par
{\obeylines\gdef\
{}}
\def\singlespace{\baselineskip=\normalbaselineskip}

\def\oneandahalfspace{\baselineskip=\normalbaselineskip
  \multiply\baselineskip by 3 \divide\baselineskip by 2}
\def\doublespace{\baselineskip=\normalbaselineskip \multiply\baselineskip by 2}

\newcount\firstpageno
\firstpageno=2
\footline={\ifnum\pageno<\firstpageno{\hfil}\else{\hfil\twelverm\folio\hfil}\fi}
\let\rawfootnote=\footnote		
\def\footnote#1#2{{\rm\singlespace\parindent=0pt\rawfootnote{#1}{#2}}}
\def\raggedcenter{\leftskip=4em plus 12em \rightskip=\leftskip
  \parindent=0pt \parfillskip=0pt \spaceskip=.3333em \xspaceskip=.5em
  \pretolerance=9999 \tolerance=9999
  \hyphenpenalty=9999 \exhyphenpenalty=9999 }
\def\dateline{\rightline{\ifcase\month\or
  January\or February\or March\or April\or May\or June\or
  July\or August\or September\or October\or November\or December\fi
  \space\number\year}}
\def\received{\vskip 3pt plus 0.2fill
 \centerline{\sl (Received\space\ifcase\month\or
  January\or February\or March\or April\or May\or June\or
  July\or August\or September\or October\or November\or December\fi
  \qquad, \number\year)}}


\hsize=6.5truein
\vsize=8.9truein
\parskip=\medskipamount
\twelvepoint		
\doublespace		
\overfullrule=0pt	



\def\title			
  {\null\vskip 3pt plus 0.2fill
   \beginlinemode \doublespace \raggedcenter \bf}

\def\author			
  {\vskip 3pt plus 0.2fill \beginlinemode
   \singlespace \raggedcenter}

\def\affil			
  {\vskip 3pt plus 0.1fill \beginlinemode
   \oneandahalfspace \raggedcenter \sl}

\def\abstract			
  {\vskip 3pt plus 0.3fill \beginparmode
   \doublespace \narrower ABSTRACT: }

\def\endtitlepage		
  {\endpage			
   \body}

\def\body			
  {\beginparmode}		

\def\head#1{			
  \filbreak\vskip 0.5truein	
  {\immediate\write16{#1}
   \raggedcenter \uppercase{#1}\par}
   \nobreak\vskip 0.25truein\nobreak}

\def\refto#1{$^{#1}$}		

\def\references			
  {\head{References}		
   \beginparmode
   \frenchspacing \parindent=0pt \leftskip=1truecm
   \parskip=8pt plus 3pt \everypar{\hangindent=\parindent}}

\gdef\refis#1{\indent\hbox to 0pt{\hss#1.~}}	

\gdef\journal#1, #2, #3, 1#4#5#6{		
    {\sl #1~}{\bf #2}, #3, (1#4#5#6)}		

\gdef\journ2 #1, #2, #3, 1#4#5#6{		
    {\sl #1~}{\bf #2}: #3, (1#4#5#6)}		

\def\refstylenp{		
  \gdef\refto##1{ [##1]}				
  \gdef\refis##1{\indent\hbox to 0pt{\hss##1)~}}	
  \gdef\journal##1, ##2, ##3, ##4 {			
     {\sl ##1~}{\bf ##2~}(##3) ##4 }}

\def\refstyleprnp{		
  \gdef\refto##1{ [##1]}				
  \gdef\refis##1{\indent\hbox to 0pt{\hss##1)~}}	
  \gdef\journal##1, ##2, ##3, 1##4##5##6{		
    {\sl ##1~}{\bf ##2~}(1##4##5##6) ##3}}

\def\prd{\journal Phys. Rev. D, }

\def\prl{\journal Phys. Rev. Lett., }

\def\np{\journal Nucl. Phys., }

\def\pl{\journal Phys. Lett., }

\def\endreferences{\body}

\def\figurecaptions		
  {\endpage
   \beginparmode
   \head{Figure Captions}
}

\def\endpage			
  {\vfill\eject}

\def\endpaper			
  {\endmode\vfill\supereject}


\def\ref#1{Ref. #1}			
\def\Ref#1{Ref. #1}			

\def\frac#1#2{{\textstyle #1 \over \textstyle #2}}

\def\sla{\raise.15ex\hbox{$/$}\kern-.57em}
\def\leaderfill{\leaders\hbox to 1em{\hss.\hss}\hfill}
\def\twiddle{\lower.9ex\rlap{$\kern-.1em\scriptstyle\sim$}}
\def\bigtwiddle{\lower1.ex\rlap{$\sim$}}
\def\gtwid{\mathrel{\raise.3ex\hbox{$>$\kern-.75em\lower1ex\hbox{$\sim$}}}}
\def\ltwid{\mathrel{\raise.3ex\hbox{$<$\kern-.75em\lower1ex\hbox{$\sim$}}}}
\def\square{\kern1pt\vbox{\hrule height 1.2pt\hbox{\vrule width 1.2pt\hskip 3pt
   \vbox{\vskip 6pt}\hskip 3pt\vrule width 0.6pt}\hrule height 0.6pt}\kern1pt}

\refstylenp

\catcode`@=11
\newcount\r@fcount \r@fcount=0
\newcount\r@fcurr
\immediate\newwrite\reffile
\newif\ifr@ffile\r@ffilefalse
\def\w@rnwrite#1{\ifr@ffile\immediate\write\reffile{#1}\fi\message{#1}}

\def\writer@f#1>>{}
\def\referencefile{
  \r@ffiletrue\immediate\openout\reffile=\jobname.ref%
  \def\writer@f##1>>{\ifr@ffile\immediate\write\reffile%
    {\noexpand\refis{##1} = \csname r@fnum##1\endcsname = %
     \expandafter\expandafter\expandafter\strip@t\expandafter%
     \meaning\csname r@ftext\csname r@fnum##1\endcsname\endcsname}\fi}%
  \def\strip@t##1>>{}}

\def\citeall#1{\xdef#1##1{#1{\noexpand\cite{##1}}}}
\def\cite#1{\each@rg\citer@nge{#1}}	

\def\each@rg#1#2{{\let\thecsname=#1\expandafter\first@rg#2,\end,}}
\def\first@rg#1,{\thecsname{#1}\apply@rg}	
\def\apply@rg#1,{\ifx\end#1\let\next=\relax
\else,\thecsname{#1}\let\next=\apply@rg\fi\next}

\def\citer@nge#1{\citedor@nge#1-\end-}	
\def\citer@ngeat#1\end-{#1}
\def\citedor@nge#1-#2-{\ifx\end#2\r@featspace#1 
  \else\citel@@p{#1}{#2}\citer@ngeat\fi}	
\def\citel@@p#1#2{\ifnum#1>#2{\errmessage{Reference range #1-#2\space is bad.}%
    \errhelp{If you cite a series of references by the notation M-N, then M and
    N must be integers, and N must be greater than or equal to M.}}\else%
 {\count0=#1\count1=#2\advance\count1 by1\relax\expandafter\r@fcite\the\count0,%
  \loop\advance\count0 by1\relax
    \ifnum\count0<\count1,\expandafter\r@fcite\the\count0,%
  \repeat}\fi}

\def\r@featspace#1#2 {\r@fcite#1#2,}	
\def\r@fcite#1,{\ifuncit@d{#1}
    \newr@f{#1}%
    \expandafter\gdef\csname r@ftext\number\r@fcount\endcsname%
                     {\message{Reference #1 to be supplied.}%
                      \writer@f#1>>#1 to be supplied.\par}%
 \fi%
 \csname r@fnum#1\endcsname}
\def\ifuncit@d#1{\expandafter\ifx\csname r@fnum#1\endcsname\relax}%
\def\newr@f#1{\global\advance\r@fcount by1%
    \expandafter\xdef\csname r@fnum#1\endcsname{\number\r@fcount}}

\let\r@fis=\refis			
\def\refis#1#2#3\par{\ifuncit@d{#1}
   \newr@f{#1}%
   \w@rnwrite{Reference #1=\number\r@fcount\space is not cited up to now.}\fi%
  \expandafter\gdef\csname r@ftext\csname r@fnum#1\endcsname\endcsname%
  {\writer@f#1>>#2#3\par}}

\def\ignoreuncited{
   \def\refis##1##2##3\par{\ifuncit@d{##1}%
     \else\expandafter\gdef\csname r@ftext\csname r@fnum##1\endcsname\endcsname%
     {\writer@f##1>>##2##3\par}\fi}}

\def\r@ferr{\endreferences\errmessage{I was expecting to see
\noexpand\endreferences before now;  I have inserted it here.}}
\let\r@ferences=\references
\def\references{\r@ferences\def\endmode{\r@ferr\par\endgroup}}

\let\endr@ferences=\endreferences
\def\endreferences{\r@fcurr=0
  {\loop\ifnum\r@fcurr<\r@fcount
    \advance\r@fcurr by 1\relax\expandafter\r@fis\expandafter{\number\r@fcurr}%
    \csname r@ftext\number\r@fcurr\endcsname%
  \repeat}\gdef\r@ferr{}\endr@ferences}


\let\r@fend=\endpaper\gdef\endpaper{\ifr@ffile
\immediate\write16{Cross References written on []\jobname.REF.}\fi\r@fend}

\catcode`@=12

\citeall\refto		
\citeall\ref		%
\citeall\Ref		%

\ignoreuncited
\referencefile

\def\L{\Lambda}
\def\th{\theta}
\def\r{\refto}

\def\sla{\raise.15ex\hbox{$/$}\kern-.57em}
\def\gev{{\rm \,Ge\kern-0.125em V}}
\def\mev{{\rm \,Me\kern-0.125em V}}
\def\bfk{{\bf k}}

\singlespace

\rightline{ACT--09/96}
\rightline{TAC--96--015}
\rightline{UCSBTH--96--19}
\rightline{UM--AC--96--05}
\rightline{October 1996}
\rightline{hep-ph/9610405}

\title
BARYOGENESIS DURING REHEATING IN NATURAL INFLATION
AND COMMENTS ON SPONTANEOUS BARYOGENESIS

\author Alexandre Dolgov$^{*}$
{\rm Teoretisk Astrofysik Center\\
 Juliane Maries Vej 30, DK-2100, Copenhagen, Denmark $^{1)}$
\footnote
{1)Also: ITEP, Bol. Cheremushkinskaya 25, Moscow 113259, Russia.\qquad\qquad\qquad\qquad\qquad}}

\author Katherine Freese$^{\dagger}$
{\rm Physics Department, University of Michigan, Ann Arbor, MI 48109}
\author Raghavan Rangarajan$^{\triangle}$
{\rm Astroparticle Physics Group, Houston Advanced Research Center\\
The Mitchell Campus, The Woodlands, TX 77381}
\author Mark Srednicki$^{\diamond}$
{\rm Department of Physics, University of California, Santa Barbara, CA
93106}

\abstract
\baselineskip16pt
We calculate the baryon asymmetry created by the decay of
a pseudo Nambu-Goldstone boson whose interactions violate baryon number
conservation.  Our results are in disagreement with previous results in the
original spontaneous baryogenesis models for the asymmetry produced by
the decay of an oscillating scalar field with B number violating derivative
couplings;  we find that the net baryon number density is proportional
to $\th_i^3$, where $\th_i$ is the amplitude of the PNGB-field in
natural inflation at the onset of reheating.  We also discuss our
disagreement with the interpretation of $\dot\theta$ as an effective
chemical potential for baryon number in spontaneous baryogenesis models.
While our calculation of the asymmetry is carried out in the context of
natural inflation our approach is generally valid for baryogenesis
models using decaying classical fields.  In the Appendices, we include a
complete derivation of the number density of particles produced by the
decay of a classical scalar field; this number density is proportional
to the integral over momenta of the one pair production amplitude.
 
${}^*$dolgov@tac.dk  \hfill
${}^\dagger$ktfreese@umich.edu

${}^\triangle$raghu@diana.tdl.harc.edu \hfill
${}^\diamond$mark@tpau.physics.ucsb.edu

\centerline {\it Submitted to Phys. Rev. D }

\endtitlepage

\body
\baselineskip=17.5pt

{\bf Section 1: Introduction}

In this paper, 1) we calculate the baryon asymmetry obtained during
reheating following natural inflation using an approach
that is generally valid for baryogenesis
models using decaying classical fields.  Our results are in disagreement
with the
the results presented in the original
spontaneous baryogenesis papers.
2)  We discuss an objection to the effective chemical potential
interpretation used in models of spontaneous baryogenesis.

In natural inflation the role of the inflaton is played by a pseudo
Nambu-Goldstone boson, hereafter referred to as $\th$, with a potential
of the form\r{freeseetal90}
$$V(\th)=\L^4(1- \cos\th)\,.\eqno(1.1)$$
This model was proposed to ``naturally" provide the 
flat potential required for inflation to work \r{steinhardtturner84},
\r{adamsfreeseguth91}.
Here $\theta = \Phi/f$, where $\Phi$ is a complex scalar field
and $f$ is the scale at which a global symmetry is spontaneously
broken; soft explicit symmetry breaking takes place at a lower
scale $\Lambda$.  From eq.~(1.1) one can see that the height
of the potential is $2 \Lambda^4$ while the width is $f$.
Since the scales of spontaneous and explicit symmetry
breaking can ``naturally" be separated by several orders
of magnitude, one can obtain $\Lambda \leq 10^{-3} f$
as required for successful 
inflation\r{models}.
$\vphantom{X\r{changetal85,hillross88,friemanetal92,deribnil,dinerosewit
,krasnikov87,caslalmunross90,KDLP,ovrutthomas9192}}$

In ref.\r{adamsetal} an extensive study of the conditions under which
the $\th$ field can drive inflation has been obtained.  After the period
of inflation, the energy density of the $\th$ field is converted to
radiation during reheating through its decay to other forms of matter
as it oscillates in its potential.  Below we shall assume that $\th$ is
coupled only to fermions.  We treat $\th$ as a classical scalar field
coupled to quantized fermion fields $Q$ and $L$ via an interaction term 
of the form $\bar Q L e^{i\theta} + \bar L Q e^{-i\theta}$,
where $Q$ carries baryon number but $L$ does not.  
We show that the decay of $\th$ gives rise to a net baryon number
density $(n_b-n_{\bar b})$
proportional to $\th_i^3$, where $\th_i$ is the value of the $\th$ field
at the onset of reheating.  

Our result disagrees with the calculation in
the original spontaneous baryogenesis papers\r{cohkap8788} where it was
argued that the asymmetry is proportional to $\th_i$ to the first power,
independent of the details of the baryon number violating couplings of the
$\theta$ field.  Specifically, in 
previous work, Cohen and Kaplan\r{cohkap8788} 
considered any theory in which
a scalar field is derivatively coupled to the baryon current $J^\mu$
with a term in the interaction Lagrangian of the form
${\cal L}_{\rm int} \propto \partial_\mu \theta J^\mu$, and
derived an expression for the baryon asymmetry produced by the
decay of the scalar field as it oscillates about its minimum.
The pseudo Nambu-Goldstone boson in natural inflation can serve as an
example of such a scalar field. 
Cohen and Kaplan obtained 
$|\dot n_B| = \Gamma f^2 |\dot \theta|$, where $\Gamma$
is the decay rate of the $\theta$ field and
$n_B$ is the net baryon number density.  This gives
$$
|\Delta n_B| = \Gamma f^2 |\Delta \theta| \, .
\eqno (1.2)$$
Below we discuss our concerns with this conclusion
and present calculations for the specific case of
eq.~(1.1); our results {\it disagree} with eq.~(1.2).
We also comment on our objections to interpreting $\dot\theta$
as a chemical potential when $\ddot\theta$ is small,
as was done in ref.\r{cohkap8788};
we argue that a Lagrangian term $\dot\theta J^0$ 
does not appear in the Hamiltonian, and therefore
it is incorrect to identify $\dot\theta$ with an effective chemical
potential for baryon number.

The framework of this paper is as follows.  In Section 2, we write down
the Lagrangian density for the inflaton field and present the equation
of motion for $\th$ as it oscillates during the reheating phase, as
derived in ref.\r{dolgovfreese}.  In Section 3 we discuss our concerns with
eq.~(1.2) as obtained in ref.\r{cohkap8788} 
(these concerns were raised in an earlier paper\r{dolgovfreese} 
by two of the authors [Dolgov and Freese]).  We then proceed to 
calculate the total baryon number and antibaryon number produced during
the decay of the $\th$ field, and find a net baryon number
density $(n_b - n_{\bar b})$ proportional to $\theta_i^3$.
We also show that the energy density of the
produced particles is equal to the initial energy density of the $\th$
field as a a check on our calculation.  In Section 4, we 
discuss how constraints on parameters in natural inflation
obtained in ref.\r{adamsetal} affect the quantitative results for
baryogenesis.
We also discuss our objections to the thermodynamic generation of the
baryon asymmetry via an effective chemical potential interpretation in
models of spontaneous baryogenesis. Finally we summarize our results.
In the Appendices we provide details of the calculations outlined in the
main body of the paper.
In particular, in Appendices A and B, we include derivations of the number
density of particles produced by the decay of a classical
scalar field; the number density of particles produced is proportional
to the integral over momenta of the one pair production amplitude.

{\bf Section 2:  The Model}

As in ref.\r{dolgovfreese} we consider a simple model involving a complex
scalar field $\Phi$ and fermion fields $Q$ and $L$ with the Lagrangian
density
\footnote
{$^1$}{
We use a metric (--1,1,1,1).
}
$${\cal L} = -\partial_\mu \Phi^*
\partial^\mu \Phi - V(\Phi^* \Phi) + i\bar Q \gamma^\mu \partial_\mu Q
+ i \bar L \gamma^\mu \partial_\mu L 
-m_Q \bar Q Q - m_L \bar L L
+ (g \Phi \bar Q L + \rm{h.c.}) \, .
\eqno(2.1)$$
Note that, despite their names,  $Q$ and $L$ cannot be actual quarks
and leptons, since the interaction term does not conserve color.
They could, however, represent heavy fermions with other interactions
with the fields of the Standard Model which fix the assignments
of global charges.  In particular, we shall assume that the field $Q$
carries baryon number while the field $L$ does not.
The U(1) symmetry that corresponds to baryon number is therefore
identified as
$$\Phi \rightarrow e^{i \alpha} \Phi \, , \qquad\qquad
  Q \rightarrow e^{i \alpha} Q \, , \qquad\qquad
  L \rightarrow L \, . \eqno(2.2)$$
We assume that this global symmetry is spontaneously broken at an
energy scale $f$ via a potential of the form
$$ V(|\Phi|) = \lambda \bigl(\Phi^* \Phi - f^2/2 \bigr)^2 \, .
\eqno(2.3)$$
The resulting scalar field vacuum expectation value (VEV)
is $\langle \Phi \rangle = f e^{i\phi/f}/\sqrt{2}$.

Below the scale $f$, we can neglect the radial mode of
$\Phi$ since it is so massive that it is frozen out;  
$m_{\rm radial} = \lambda^{1/2} f$.
The remaining light degree of freedom
is $\phi$,
the Goldstone boson of the
spontaneously broken $U(1)$.  
For simplicity of notation we introduce the dimensionless angular field
$\theta\equiv\phi/f$.  We then obtain
an effective Lagrangian density for $\theta$, $Q$, and $L$ of the form
$${\cal L}_{\rm eff}
= -{f^2\over 2} \partial_\mu \theta \partial^{\mu} \theta + i \bar Q
\gamma^{\mu} \partial_\mu Q + i \bar L
\gamma^{\mu} \partial_\mu L 
-m_Q \bar Q Q -m_L \bar L L
+\bigl({g\over\sqrt{2}} f\bar Q L e^{i\theta} + h.c.\bigr)
\, . \eqno(2.4)
$$

The global symmetry is now realized in the Goldstone mode:
${\cal L}_{\rm eff}$ is invariant under
$$Q \rightarrow e^{i \alpha} Q\, , \qquad\qquad L \rightarrow L \, , 
\qquad\qquad \theta \rightarrow
\theta + \alpha \, . \eqno(2.5)$$  With a rotation of the form
in eq.~(2.5) with $\alpha = - \theta$, the 
Lagrangian can alternatively be written
$${\cal L}_{eff}
= -{f^2\over 2} \partial_\mu \theta \partial^{\mu} \theta + i \bar Q
\gamma^{\mu} \partial_\mu Q + i \bar L
\gamma^{\mu} \partial_\mu L 
-m_Q \bar Q Q -m_L \bar L L
+ ({g\over\sqrt{2}} f\bar Q L + h.c.) + \partial_\mu
\theta J^\mu \;, \eqno(2.6)
$$
where the fermion current derives from the $U(1)$ symmetry;
here, $J^\mu = \bar Q \gamma^\mu Q$.  

We now assume that the symmetry (2.2) is also subject to a small explicit
breaking which gives rise to 
a potential as in eq.~(1.1) 
and which provides a nonzero mass for the field $\theta$.
This explicit symmetry breaking could come from Planck scale physics.
Alternatively, one can imagine a scenario similar to that involving the 
QCD axion where, at energy scales of the order of $\L_{\rm QCD}$, instanton 
effects create the fermion condensate 
$\langle\bar \psi \psi \rangle \sim \Lambda_{\rm QCD}^3$, giving rise
to a mass term for the axion.  
Note that
for the natural inflation model, the required mass
scales are much higher than for the QCD axion.  The width
of the potential must be roughly the Planck mass in order
to achieve enough e-foldings of inflation, and the 
height of the potential must be roughly
$\Lambda^4 \sim [10^{16}$ GeV]$^4$ in order for
density perturbations appropriate for structure
formation to be produced (see the Discussion section
at the end of the paper for more detail).  Consequently the scale
at which the relevant gauge group ({\it not} QCD)
must become strong is roughly the GUT scale.
These and other mechanisms such as those 
found in technicolor and schizon models for generating a potential for 
pseudo Nambu-Goldstone bosons are discussed 
in ref.\r{models,adamsetal}.

Initially, as the $\th$ field rolls down towards the minimum of its
potential, its potential energy drives inflation.  
Let $\th_i$ be the value of the $\th$ field at the 
beginning of the reheating epoch, after inflationary expansion has ended.
(We shall ignore spatial variations in
the $\theta$ field.)
During the reheating epoch the $\th$
field oscillates about the minimum of its potential.  While $\th$
oscillates it decays to the fields $Q$ and $L$.  The interactions of the
fermionic fields create a thermal bath thereby reheating the universe.
Note that we must take $g \ll 1$
so that fermion masses generated for the fermions from
the Yukawa coupling, $m_\psi \sim g f$, are small enough that
the fermions can in fact be produced by decays of the
pseudo Nambu-Goldstone bosons.  See ref. [10] in Dolgov and 
Freese\r{dolgovfreese}
for further discussion of this point.

The equation of motion for the $\th$ field with the back reaction
of the produced fermions
was rigorously derived in the one loop approximation in ref.\r{dolgovfreese}. 
For small deviations of $\theta$ from the equilibrium the potential can
be approximated as $V(\theta)={1\over2}m_R^2 f^2 \theta^2$ and 
the equation of motion
during the oscillating phase
can be effectively written in the well known form
$$\ddot \theta + m_R^2 \theta + \Gamma \dot \theta = 0 \, , \eqno(2.7)$$ 
where $m_R$ is the renormalised $\th$ mass defined as 
$\lim_{w\to\infty}m_R^2\bigl[1+{g^2\over4\pi^2}\log(2w/m_R)\bigr]=m^2$,
where $m$ is the bare mass of the $\theta$ field, and
$\Gamma \equiv g^2 m_R /8\pi$.  (Our expressions above differ by a
factor of 2 from those in ref.\r{dolgovfreese} because a 
factor of $1/ \sqrt 2$ was dropped from eq.~(2.5) in
ref.\r{dolgovfreese}.)  The solution to this equation is
$$
\theta(t) = \theta_i e^{- \Gamma t /2} \cos(m_R t)\, .
\eqno(2.8)$$
where we have assumed that the initial velocity of the $\th$ field is
negligible and have therefore set an arbitrary phase
in the cosine to zero.  The results obtained
below can be easily generalized for arbitrary initial conditions.
The above solution was derived assuming $m_Q=m_L=0$.
However, it can be shown that non-zero values
of $m_Q$ and $m_L$ will not change the
solution for $\th$ significantly as long as $m_Q, \, m_L \ll m_R$, which we
shall assume below.

{\bf Section 3:  Baryogenesis}

{\it Previous Calculations and Concerns:}
In previous work, Cohen and Kaplan\r{cohkap8788} considered any theory in which
a scalar field is derivatively coupled to the baryon current with a term
in the interaction Lagrangian of the form
${\cal L}_{int} \propto \partial_\mu \theta J^\mu$, and
derived an expression for the baryon asymmetry produced by the
decay of the scalar field as it oscillates about its minimum.  From 
eq.~(2.6) one can see that our pseudo Nambu-Goldstone boson is an
example of such a scalar field as it has the appropriate
coupling.  Cohen and Kaplan obtained 
$|\dot n_B| = \Gamma f^2 |\dot \theta|$,
where $n_B$ is the net baryon number density.  This gives
$$
|\Delta n_B| = \Gamma f^2 |\Delta \theta| \, .
\eqno (3.1)
$$
In a previous paper\r{dolgovfreese} by two of the authors [Dolgov and Freese],
several concerns with this interpretation were raised.
We will outline two of these concerns again here, and then 
proceed with a direct calculation of the baryon asymmetry.
Our results will {\it disagree} with eq.~(3.1).

One concern is as follows:  in making the identification
$|\dot n_B| = \Gamma f^2 |\dot \theta|$,
one is comparing an operator equation, namely, the Euler-Lagrange
equation
$\ddot \theta + m^2 \theta = \dot n_B/f^2$, with an equation of the form
of eq.~(2.7) which is obtained after vacuum averaging.  
In ref.\r{dolgovfreese}
the average value $\langle \dot n_B \rangle$
was found to be not just $-\Gamma f^2 \dot \theta$ but a more
complicated expression (eq.~(3.3) in ref.\r{dolgovfreese}).  

A second concern is with regard to energy conservation.
The initial energy density of the field $\theta$ which creates the
baryons and antibaryons is $\rho_\theta(t_i) \sim f^2 m^2 \theta_i^2$.
At the end some of this energy density has been converted to
baryons and antibaryons, with energy density 
$\rho(t_f)>n_B E_B$ where
$E_B \sim m$ is the characteristic
energy of the produced fermions (note that $n_B$ refers to the difference
between baryon and antibaryon number densities
and not to the total number density of produced
particles).  Clearly it must be
true that $\rho(t_f) < \rho_\theta (t_i)$.  If we
were to use eq.~(3.1) we would see that this requires 
$\Gamma < \Delta \theta m$. From the definition of $\Gamma$
we see that this is satisfied for small values of 
coupling constant $g$ as long as $\Delta \theta$ is
not too small; for small values of $\Delta \theta$,
this relationship can never be satisfied.

{\it New Calculations and Results:}
We now proceed to calculate the net baryon number density of the particles
produced during reheating.  We perform an explicit calculation
and find a different result from eq.~(3.1).
The $\th$ field decays to either
$Q \bar L$ pairs or $\bar Q L$ pairs.  (The $Q$ and $L$ fields are not
the mass eigenstates.  Later in this section we consider effects of
oscillations between $Q$ and $L$ fields.)  As mentioned earlier, we
treat the $\th$ field classically, $Q$ and $L$ are quantum fields and
$Q$ carries baryon number.  For now we ignore any dilution of the baryon
number density due to the expansion of the universe.

As shown in Appendix A
with the Bogolyubov transformation method\r{bogolyubov},
the average number density $n$ of particle antiparticle pairs
produced by decay of a homogeneous classical scalar field,
to lowest order in perturbation theory,
 is given by
$$n={1\over V} \sum_{s_1,s_2} \int \widetilde{dp_1} \; \widetilde{dp_2} \;
|A|^2 \, , \eqno(3.2)$$
where $A$ is the one pair production amplitude, subscripts
1 and 2 refer to the final particles produced and  
$\widetilde{dp} = d^3 p/[(2\pi)^3 2p^0]$.
Eq.~(3.2) can also be obtained using the method
presented in Sec. 4-1-1 of
ref.\r{itzyksonzuber80}, as discussed in Appendix B.

Thus, to lowest order in perturbation theory,
the average number density of $Q \bar L$ pairs produced during reheating
in our model is given by
\footnote
{$^2$}{
Throughout the paper, a state 
$\langle A(p_1,s_1),\bar B(p_2,s_2)|$ corresponds to a final state with
an $A$ particle of momentum $p_1$ and spin $s_1$ and an anti-$B$ particle
with momentum $p_2$ and spin $s_2$.
}
$$
n(Q, \bar L)={1\over V} \sum_{s_Q,s_{\bar L}} \int \widetilde{dp} \; 
\widetilde{dq} \;
\bigl|\langle  Q(p,s_Q), \bar L(q,s_L)|0\rangle\bigr|^2 \, . \eqno(3.3)$$
We take
$$Q = \sum_s \int \widetilde{dk}\;
\bigl[u_k^s b_k^s e^{+i k \cdot x} + v_k^s d_k^{s\dagger} e^{-i k \cdot x}
\bigr] \, \eqno(3.4)$$
and a similar expression for $L$.
Here $\{b_k^s,b_{k^\prime}^{s^\prime\dagger}\}
=\{d_k^s,d_{k^\prime}^{s^\prime\dagger}\}
=(2\pi)^3 2k^0\delta^3({\bf k-k^\prime})\delta_{ss^\prime}$.
Standard algebra gives 
$$\eqalign{n(Q,\bar L) &=
{1\over V}\sum_{s_Q,s_{\bar L}}\int \widetilde{dp}\; \widetilde{dq}\;\biggl|
\langle Q(p,s_Q),\bar L(q,s_L)|i{g\over\sqrt 2}\int d^4x  \, \bar Q(x) L(x)
e^{i \theta(x)} |0\rangle \biggr|^2\cr
&= {g^2 f^2\over 2V} \int \widetilde{dp} \; \widetilde{dq} \;
\biggl| (2\pi)^3 \delta^3({\bf p+q})
\int^{\infty}_{-\infty} dt \, e^{i 2\omega t + i \th(t)} \biggr|^2 
\mathop{\rm Tr}[(-\sla p +m_Q)(-\sla q -m_L)]
\,  \cr} \eqno(3.5)$$
where $2\omega=p^0+q^0$.
We obtain a similar expression for
$n(L,\bar Q)$ with $\th(t)$ replaced by $-\th(t)$.
We set the baryon number density $n_b$ to be equal to
$n(Q, \bar L)$ and the antibaryon number density $n_{\bar b}$ to be
equal to $n(L,\bar Q)$.  Then we have 
$$n_{b, \bar b} =
{g^2 f^2 \over 2 \pi^2} \int d\omega \, \omega^2 \,
\biggl|\int_{-\infty}^{\infty} 
dt\,e^{2 i \omega t}  e^{\pm i \theta(t)}\biggr|^2 \, , \eqno(3.6)$$
where the $+$ sign in the exponent refers to baryon number
and the $-$ sign to antibaryon number.
To carry out the integral over time we expand $e^{i \th}$ as
$$1+i\th-\th^2/2 \, , \eqno(3.7)$$ 
valid for small $\th$, and use
$$ \th (t) = \cases{\th_i & for $t\le0$ \cr
                \th_i e^{-\Gamma t/2} \cos(m_Rt) & for $t\ge 0\, .$ \cr}
\eqno(3.8)
$$
We also use a convergence factor at early times to regularize the
integral.  
We will examine a series of possible terms to find the
first nonzero contribution in perturbation theory.
The lowest order term comes from using $e^{i \theta} = 1$
from eq.~(3.7) in eq.~(3.6) 
and gives $\int dt e^{2i\omega t} \propto \delta(2\omega)
=0$ since we can not have $\omega = 0$ for particle production.
The next term in the expansion, the $\theta$ term in (3.7),
when squared gives the same contribution to $n_b$ and to
$n_{\bar b}$.  In order to obtain an asymmetry one must consider
cross terms.  The lowest order cross term that gives a nonzero
contribution to the baryon asymmetry is
$$n_b - n_{\bar b} = 2 \times {g^2 f^2 \over 2\pi^2}
\int d\omega \, \omega^2 \Biggl[
{\tilde \theta(2 \omega) \,
\left[\widetilde{\theta^2}(2 \omega)\right]^* \over 2i}
\, + {\rm h.c.}\Biggr], \eqno(3.9)$$
where $h.c.$ refers to hermitian conjugate,
$$\tilde \theta (2 \omega) =  \int_{-\infty}^{\infty} dt \, e^{2 i \omega t} \,
\theta (t) \, \eqno(3.10a)$$
and
$$\widetilde{\theta^2}(2\omega) =  \int_{-\infty}^{\infty} dt \, e^{2i\omega t} 
\, \theta^2(t) \, . \eqno(3.10b)$$  
The factor of 2 in eq.~(3.9) arises 
from the fact that the cross terms in $n_b$ and 
$n_{\bar b}$ terms are the same up to a minus sign.
One can see from the form of eq.~(3.9) that
we expect the asymmetry to be proportional to $\theta^3$.
The details of this calculation are outlined in Appendix C,
and the results are presented here.

We obtain
$$n_b={1\over4}m_Rf^2\th_i^2 +{g^2\over 32\pi} m_Rf^2\th_i^3\, \eqno(3.11)$$
$$n_{\bar b}={1\over4}m_Rf^2\th_i^2
-{g^2\over 32\pi} m_Rf^2\th_i^3\, \eqno(3.12)$$
Therefore,
$$\eqalign{
n_B\equiv n_b-n_{\bar b}&={g^2\over 16\pi} m_Rf^2\th_i^3 \cr
&={1\over 2}\Gamma f^2 \th_i^3
\cr}\eqno(3.13)
$$

We notice that the net baryon number density is proportional to $\th_i^3$.  This
disagrees with the calculation in ref.\r{cohkap8788} which gives an
asymmetry proportional to $\th_i$.  We also note that
the number density of pairs of particles $n_b+n_{\bar b}$ is equal to
${1\over2} m_R f^2\th_i^2$.  Since the energy per pair of particles is
$m_R$, the
energy density in the produced particles is ${1\over2} m_R^2f^2\th_i^2$, which
agrees with the initial energy density of the $\th$ field.  We have also
done the calculation of
$$
\rho_{\rm final}={1\over V}
\sum_{s_Q,s_L} \int \widetilde{dp} \; \widetilde{dq} \; \bigl(p^0+q^0\bigr)
\bigl[ |\langle  Q(p,s_Q),\bar L(q,s_L) |0\rangle|^2 +
       |\langle  L(q,s_L),\bar Q(p,s_Q) |0\rangle|^2 \bigr]
\,\eqno(3.14)$$
and have verified that we obtain ${1\over2} m_R^2f^2\th_i^2$.

{\it Mass Mixing:}  
In many cases
eq.~(3.13) is not yet the complete story because of mass mixing.
As we mentioned earlier the $Q$ and $L$ fields are not mass eigenstates.
Therefore a particle which is produced as a $Q$ may later 
rotate into an $L$.  This effect must be taken into account.
Eq.~(3.13) is completely correct for the case where the
fermions $Q$ and $L$ are converted immediately to regular
quarks $q$ and leptons $l$ as soon as they are produced
(assuming that the temperature is low enough that
the $q$ and $l$ cannot convert back into $Q$ and $L$).
In that case, there is no opportunity for mixing to 
take place, e.g., there is no opportunity for $Q$ to
convert to an $L$.  On the other hand, if $Q$ and $L$
do not decay immediately into stable lighter mass particles
with appropriate quark quantum numbers, they may have
the chance to mix into one another.
One can calculate the effects of mixing in either the
$Q,L$ basis or in the basis of mass eigenstates;
below we will do both.

The mass matrix in the $(Q,L)$ basis is
$$\left(\matrix{m_Q&-gf/\sqrt 2\cr
-gf/\sqrt 2&m_L\cr}\right)\,.\eqno(3.15)$$
The mass eigenstates are
$$\psi_1={L+\epsilon Q \over \sqrt{1+\epsilon^2}}\quad{\rm and}\quad
\psi_2={Q-\epsilon L \over \sqrt{1+\epsilon^2}}\,\eqno(3.16)
$$
with masses $m_Q-gf/(\sqrt{2}\epsilon)$ and
$m_L+gf/(\sqrt{2}\epsilon)$ respectively, where
$\epsilon=\sqrt{2}gf/(\Delta m +\sqrt{(\Delta m)^2 +2 g^2 f^2})$ and
$\Delta m=m_Q-m_L$.  
Note that $\Delta m = 0$ corresponds to $\epsilon = 1$.

In the $\psi_1, \psi_2$ basis,
one can now calculate the baryon asymmetry as
a sum of terms, each of which is a product of a number density of
produced particle/antiparticle pairs times the (time averaged) quark
content of the pair,
$$
n_B=n(\psi_1,\bar\psi_2)|\langle Q|\psi_1\rangle|^2
+n(\psi_2,\bar\psi_1)|\langle Q|\psi_2\rangle|^2
-n(\psi_1,\bar\psi_2)|\langle \bar Q|\bar\psi_2\rangle|^2
-n(\psi_2,\bar\psi_1)|\langle \bar Q|\bar\psi_1\rangle|^2
\, .\eqno(3.17)
$$
Here $n(\psi_1,\bar\psi_2)$ and $n(\psi_2,\bar \psi_1)$ are the number
densities of $\psi_1$ and $\bar \psi_2$ pairs and
$\psi_2$ and $\bar \psi_1$ pairs respectively; and
$|\langle Q|\psi_i\rangle|^2$ is the probability
that a particle which is produced as a $\psi_i$ (where
$i = 1,2$) is measured
as a $Q$.  Hence, for example,
the first term is the product of the number density of $\psi_1$
$\bar \psi_2$ pairs produced times the quark content
of $\psi_1$.  

Note that we are here computing a time averaged
baryon asymmetry;
actually the value of the baryon asymmetry oscillates
in time, as discussed in Appendix D.  From 
eq.~(3.16) we see that the probability that 
$\psi_{1,2}$ is measured as a $Q$ is
$$|\langle Q|\psi_1\rangle|^2 = 
|\langle \bar Q|\bar \psi_1 \rangle|^2 =
{\epsilon^2 \over 1 + \epsilon^2}
\, \eqno(3.18)$$
and $$|\langle Q|\psi_2\rangle|^2 = 
|\langle \bar Q|\bar \psi_2 \rangle|^2 =
{1 \over 1 + \epsilon^2} \, . \eqno(3.19)$$

As in eq.~(3.2), the number densities of particle/antiparticle
pairs are obtained by squaring the production amplitudes
for the pairs, 
$$n_{i\bar j} = {1 \over V} \sum_{s_i,s_{\bar j}}
\int \widetilde{dk_i}\; \widetilde{dk_{\bar j}} \; 
|A_{i\bar j}|^2 \, , \eqno(3.20)$$ 
where $i$ and $j$ are either
1 or 2.  The amplitude for production of a
$ \psi_i \bar \psi_j$ pair is
$$A_{i\bar j} = \langle \psi_i,\bar \psi_j |
i\int \, d^4x ( {g\over \sqrt 2}
f e^{i \theta} \bar Q L + h.c.)| 0 \rangle \, . \eqno(3.21)$$
Using eqs.~(3.18), (3.19), and (3.20), we can write eq.~(3.17)
as
$$n_B = -{1 \over V} \sum_{s_1,s_2} \int \widetilde{dk_1}\; 
\widetilde{dk_2} \; 
 \Biggl({1- \epsilon^2 \over 1 + \epsilon^2} \Biggr) \,
[|A_{1\bar 2}|^2 - |A_{2\bar 1}|^2] \, . \eqno(3.22)$$
Using 
$$\bar Q L = \Biggl({1 \over 1+\epsilon^2}\Biggr) \, [ \bar \psi_2 \psi_1
- \epsilon \bar \psi_2 \psi_2 + \epsilon \bar \psi_1 \psi_1
- \epsilon^2 \bar \psi_1 \psi_2 ] \, \eqno(3.23)$$
and its hermitian conjugate,
we calculate the relevant production amplitudes:
$$
A_{1\bar 2}  = \langle \psi_1, \bar \psi_2| i\int d^4x \, ({g\over \sqrt 2}
f e^{i \theta} \bar 
Q L + {g\over\sqrt 2}f e^{- i \theta}  \bar L Q)|0 \rangle \eqno(3.24a)$$
to find
$$A_{1\bar 2} = i {g\over\sqrt 2} f
\Biggl({1 \over 1 + \epsilon^2} \Biggr) \, \langle \psi_1,\bar \psi_2 |
\int d^4x \,(\bar \psi_1 \psi_2 e^{-i \theta} - \epsilon^2 e^{i \theta}
\bar \psi_1 \psi_2) | 0 \rangle \, . \eqno(3.24b)$$
Now the two matrix elements in eq.~(3.24b) 
are similar to the ones
we 
calculated in eq.~(3.5), with $\bar Q L$
replaced by $\bar \psi_1 \psi_2$.  Hence, we have
$$A_{1\bar2} = \Biggl( {1 \over 1+ \epsilon^2} \Biggr) \, 
(A_{L \bar Q } - \epsilon^2
A_{Q \bar L }) \, . \eqno(3.25)$$
Similarly, 
$$A_{2\bar 1} = h.c. [A_{1\bar 2}] = \Biggl( {1 \over 1 + \epsilon^2} \Biggr) \,
(-\epsilon^2 A_{L \bar Q } + A_{Q \bar L }) \, . \eqno(3.26)$$
Thus eq.~(3.22) becomes
$$\eqalign{
n_B &=  \Biggl({1 - \epsilon^2 \over 1 + \epsilon^2 } \Biggr)^2
\Biggl(\sum_{s_{ Q}, s_{\bar L}} \int \widetilde{dk_Q}\; 
\widetilde{dk_{\bar L}}\; 
|A_{Q \bar L }|^2
- \sum_{s_{L}, s_{\bar Q}} \int \widetilde{dk_L}\; \widetilde{dk_{\bar Q}} \;
|A_{L \bar Q }|^2\Biggr) \cr
&=  \Biggl({1 - \epsilon^2 \over 1 + \epsilon^2 } \Biggr)^2
\times \hbox{our  previous answer} \, . \cr} \eqno(3.27)$$
Thus we find that
$$n_B={1\over 2}\Gamma f^2 \th_i^3
\Biggl({1-\epsilon^2\over 1+\epsilon^2}\Biggr)^2
\,. \eqno(3.28)
$$
If $m_Q=m_L$, $\epsilon=1$ and the asymmetry vanishes because in this case
the net baryon number of a $(\psi_1,\bar \psi_2)$ pair or a
$(\psi_2,\bar \psi_1)$ pair is 0 and thus no baryon asymmetry is
produced.

Another derivation of eq.~(3.28) is given in Appendix D.
In the preceding paragraphs we considered particle production
and mixing in the mass eigenstate $\psi_1, \psi_2$ basis.
In Appendix D we work in the $Q, L$ basis.
We find the oscillations of the baryon asymmetry with time,
and obtain the same expression as in eq.~(3.28) for the time averaged baryon
asymmetry.

{\it Thermalization:}
After the $\th$ field has decayed into $\psi_1$ and $\psi_2$ particles,
thermal equilibrium can be established if these particles have other
interactions with each other and with other particles.  As long as one
introduces interactions such as $\chi \bar\psi_1 \psi_1$ and
$\chi \bar\psi_2 \psi_2$ as  a part of a realistic model, the number of
$\psi_1-\bar\psi_1$ particles and of $\psi_2-\bar\psi_2$ particles does
not change, thereby preserving the baryon asymmetry.  (Interactions such as
$\chi\bar\psi_1\psi_2 + h.c.$ would, however, destroy the baryon asymmetry.)
The fields $\psi_1$ and $\psi_2$ will annihilate or decay to
lighter particles which will thermalize.  If these interactions preserve the
net baryon number, then the asymmetry will survive.

{\it Quantitative Results:}
So far we have not included the effects of the expansion of the
universe.  For baryon number created when $H\leq \Gamma$,
we may neglect the expansion and directly use the results obtained
above in eq.~(3.28) 
with $\theta_i$ replaced with the value of $\theta$ at $H=\Gamma$.
Since the $\theta$ field 
dominates the 
cosmic energy density, the condition $H=\Gamma$ fixes the amplitude of
$\theta$ at that moment to be
$$
\theta_1 = \sqrt{3/4\pi}(\Gamma m_{Pl} /fm_R) \approx 0.02 g^2m_{Pl}/f
\ll 1. 
\eqno(3.29)
$$
In the early stages of reheating with $\theta > \theta_1$,
expansion of the universe must be taken into account.

The decay of the $\theta$-field produces
relativistic $\psi_{1,2}$ and $\bar\psi_{1,2}$ with  energies
$\omega \approx m_R /2$. This state is
far from thermal equilibrium (the temperature of the
thermalized plasma in eq.~(3.30) below may be
smaller than the $\psi$ masses). The rate of
thermalization depends upon the interaction strength of the fermions created
in the $\theta$ decay. It is typically higher than the decay rate
because $g\ll 1$ to ensure reasonable fermion masses.
Thermalization
could occur either through annihilation of $\psi_1$ and $\bar \psi_1$ or 
$\psi_2$ and $\bar \psi_2$ into light
particles or through their decays and subsequent elastic scattering. Assuming
that these processes are fast we can roughly estimate the reheat temperature
in the instantaneous decay approximation,
$\rho_{\rm rad}=\rho_\theta(t=\Gamma^{-1})$, as
$$
T_{\rm reh} = (90 /8\pi^3g_*)^{1/4}\sqrt{\Gamma m_{Pl}}
\approx 0.15gg_*^{-1/4} \Lambda \sqrt{m_{Pl}/f}
\eqno(3.30)
$$
where we have taken $m_R=\Lambda^2/f$.

The entropy density after thermalization is given by
$s= 4\pi^2 g_*T^3_{\rm reh}/90$. It is conserved in the comoving volume if the
expansion of the universe 
is adiabatic, in particular in the absence of first order
phase transitions as the universe cools. 
Baryonic charge density is also assumed 
to be conserved inside a comoving volume
during and after thermalization and so the baryon-to-entropy
ratio $n_B/s$ remains
constant in the course of expansion.

First we find the baryon asymmetry produced after $H \leq \Gamma$
so that expansion may be neglected (subscript $1$
refers to this case).  Using eqs.~(3.28), (3.29) and (3.30)
we find
$$
\big( {n_B \over s} \big)_1 
\approx 10^{-4} {g^5\over g_*^{1/4}}\left({m_{Pl}\over f}\right)^{3/2}
{f \over \Lambda }
\Biggl({1-\epsilon^2\over 1+\epsilon^2}\Biggr)^2 \, .
\eqno(3.31)
$$
In the models studied in ref.\r{adamsetal} $(f/m_{Pl} ) = O(1)$ and 
$f/\Lambda = 10^6-10^3$,
so to get a reasonable baryon asymmetry we need a rather large coupling,
$g>10^{-2}$ (for $\epsilon\ll 1$).

In fact the asymmetry should be noticeably larger than that given by eq.~(3.31).
The result that we got above refers to the case  when $H<\Gamma$ but the
process of particle production starts much earlier when $H\approx m_R$ and
the inflaton field begins to oscillate around the bottom of the potential.
The net baryon number density produced while $H>\Gamma$ 
is again proportional to  $\theta^3$, as it is associated with the
interference between the $\theta$ and the $\theta^2$ terms in 
$|\int dt\,e^{2i\omega t}(1+i\theta-\theta^2/2)|^2$
in eq.~(3.6).
The generation of the asymmetry is more efficient at early times $(H>\Gamma)$
since the amplitude of the $\theta$-field, which goes down with the scale 
factor as $R^{-3/2}$, is larger. 
However when $H>\Gamma$ one must include the effects of the expansion of
the universe on the production of the baryon asymmetry.
This makes the exact calculations
considerably more complicated. Still we can roughly estimate the asymmetry
in the following way. The difference between the production of particles and
antiparticles is most profound at early times, $\Delta t_a \sim 1/m_R$,
when $\theta$ is large.
The total number of particles produced in time $\Delta t_a$ is
proportional to
$\Gamma \Delta t_a n_\theta$ and, as we mention above, the baryon number
asymmetry must vary as
$\theta^3$. 
Therefore, 
a reasonable estimate of the net baryon number density 
created while $H>\Gamma$ is $n_B \sim \Gamma f^2 \theta_i^3$.
Between the time of peak production of baryon
asymmetry at $t_a \sim 1/m_R$ and the peak entropy production at
$t_b \sim 1/ \Gamma$ we will take the baryon asymmetry to
be diluted by a factor of $(R_a/R_b)^3 \sim (t_a/t_b)^2
\sim (\Gamma/m_R)^2$ due to the expansion of the universe,
where we have taken the universe to behave as matter dominated
with $R \propto t^{2/3}$ in the usual fashion during reheating.
Thus the baryon-to-entropy ratio at time $t_b$ and afterwards
is $\big( {n_B \over s} \big)_2 \sim \Gamma f^2\theta_i^3 (\Gamma/m_R)^2/s$.
The calculation of the entropy density is exactly the same as described
above eq.~(3.31), while
the baryonic charge density is larger than the $H<\Gamma$ case by a factor of
$(\theta_i/\theta_1)^3(\Gamma/m_R)^2=\theta_i/\theta_1=m_R/\Gamma
= 8 \pi / g^2 \gg1$. 
Consequently, we get that the total baryon asymmetry of the universe is
approximately equal to
$$
\big( {n_B \over s} \big)_2
= {\theta_i\over \theta_1} \big( {n_B \over s} \big)_1 
\approx 3\times 10^{-3}{g^3\over g_*^{1/4}}
\left( {m_{Pl} \over f}\right)^{3/2} {f \over \Lambda}
\Biggl({1-\epsilon^2\over 1+\epsilon^2}\Biggr)^2 \, .
\eqno(3.32)
$$
Here subscript $2$ refers to the case where expansion has
been included.
Henceforth we use eq.~(3.32) as our estimate of the baryon asymmetry produced.

{\bf Section 4:  Discussion}

In ref.\r{adamsetal}, the authors obtain constraints on the parameters
$\L$ and $f$.  The stipulation that a large fraction of the universe
after inflation have inflated by at least 60 e-foldings gives
$f\geq0.06 M_{Pl}$.  A stronger constraint
can be obtained by requiring the formation of galaxies to
take place early enough in the history of the universe;
in this way one obtains
$f\geq 0.3 M_{Pl}$.  A constraint on $\L$ is derived by using COBE
data on the density fluctuation amplitude and is plotted in
fig.~1 of ref.\r{adamsetal}; the upper bound on $\L$ 
thus obtained ranges from $10^{13}\gev$ to
$10^{16}\gev$ for $f$ between $0.3\,M_{\rm Pl}$ and $1.2\,M_{\rm Pl}$.
If one desires the density fluctuations from inflation to be responsible
for the large scale structure of our universe and hence for the
COBE anisotropy, then $\Lambda$ must be equal to the
above values rather than simply being bounded by these numbers.

If the baryon asymmetry produced above is accompanied by an equal lepton
asymmetry, so that $B-L =0$, it will be wiped out by baryon number
violating sphaleron processes unless the reheat temperature is below
$100\gev$.  The low reheat temperature condition may be a desirable
feature of our model as many inflation models have difficulty creating a
high reheat temperature.
Furthermore, we shall require that $T_{\rm reh}>10\mev$ so
that we reproduce standard nucleosynthesis.  If, in addition,
one requires the density fluctuations from inflation to
serve as the explanation for the COBE data rather than
merely being bounded by it, then $\Lambda$ is determined
as a function of $f$ as described in the previous paragraph;
then the combination of these constraints 
implies that $10^{-14}< g < 10^{-10}$ for $\Lambda$ and $f$ equal to
$10^{13}\gev$ and $0.3 M_{Pl}$ respectively,
and the asymmetry generated by the mechanism
considered above is by far below the necessary observed value.
However if $\Lambda$ is merely bounded by COBE measurements
(density fluctuations must then be generated some other way than
by the inflation), then $g$ can be much larger as can the
baryon asymmetry.  Alternatively
if a nonzero $(B-L)$ is generated, for example, if the $L$ fields
carry no lepton number, then it is not destroyed
by the electroweak processes and the coupling constant $g$ need not be so
small.

In our perturbative calculations of the number of pairs of particles
produced we have assumed that the masses of the
fermions are smaller than the mass $m_R$ of the theta-field and that
$gf<m_{Q,L}$; otherwise the 
perturbative
approach is not applicable. This implies that 
$gf < m_R = \Lambda^2/f$ or
$g<(\Lambda/f)^2$. 
In this case, the
baryon asymmetry is rather small as $\big( {n_B \over s} 
\big)_2 < 10^{-3} (\Lambda/f)^5
(m_{Pl}/f)^{1.5} < 10^{-18}$ (in obtaining this limit we have included
the simultaneous constraint on $\Lambda$ and $f$ from density fluctuation
constraints in  ref.\r{adamsetal}).
If, however, $\theta $ is not the inflaton field, as in the original version of
the spontaneous baryogenesis scenario\r{cohkap8788}, 
then the parameters $\Lambda$
and $f$ do not necessarily satisfy the above bounds and the asymmetry may be
quite large, especially if $f\ll m_{Pl}$.   In such a case, one would
have to redo the calculation of the entropy if $\theta$ does not dominate
the energy density of the universe when it decays.  A
period of inflation prior to the decay of the PNGB would also 
be required so that
$\theta$ and, consequently, the baryon asymmetry have the same sign within 
present-day domains of sizes 10 Mpc or greater.  (Existing data do not
rule out a matter symmetric universe with domains of matter and
antimatter on scales of 10 Mpc or more\r{stecker89others}.)

An interesting possibility is that
the mass of fermions is not below $m_R$ and the perturbative approach is not
applicable. The non-perturbative calculations in this case are more complicated
and will be
presented elsewhere. 

We would also like to point out an objection to the mechanism of
creating the baryon asymmetry thermodynamically, via an effective
chemical potential interpretation, as first proposed in
ref.\r{cohkap8788} and later applied to spontaneous baryogenesis models
at the electroweak phase transition\r{ewspbar}.  
The approach in ref.\r{cohkap8788} is to identify
$\dot\theta$ in the term $\partial_\mu\theta J^\mu=-\dot\theta
J^0=-\dot\theta n_B$ in eq.~(2.6) with an effective chemical potential
that biases the B violating interactions in the universe to create more
baryons than antibaryons.  
If this is the case then, as is argued in
ref.\r{cohkap8788}, the net baryon number density in
the thermal
bath is given by the expression: $n_B \sim \dot \theta T^2$. However this
can not
be true because it contradicts the equation of motion of the Goldstone
field:
$\partial^2 \theta = -\partial_\mu J_B^\mu /f^2$. 
Assuming spatial
homogeneity, this equation gives
$n_B\sim \dot \theta f^2$. (A similar contradiction is obtained using
the equation of motion for a PNGB-field in an expanding universe.)
This contradiction is resolved because 
$\dot \theta$ does not shift energies of baryons and
antibaryons and cannot be identified with a chemical potential. 
While the term $\partial_\mu J^\mu$ exists in the Lagrangian
density in eq.~(2.6), it does not in the Hamiltonian density
$$H={\partial L_{\rm eff}\over\partial\dot\phi_i}\dot\phi_i-
L_{\rm eff}(\phi_i,\dot\phi_i)\, , \eqno(4.1)
$$
where $\phi_i$ represents all the fields in the Lagrangian density\r{voloshin}.
This is similar to the interaction of a charged particle with a magnetic field,
where the energy of the particle is not changed due to the action of the
field as the force is proportional to the velocity and orthogonal to it.
Thus the term $-\dot\theta n_B$ does not appear in the Hamiltonian
density and $\dot\theta$
can not be interpreted as an effective chemical potential.

As mentioned above, 
the approach of ref.\r{cohkap8788} has been applied to create
the baryon asymmetry in spontaneous baryogenesis models at the
electroweak phase transition\r{ewspbar}.  The role of the $\theta$ field
is associated with the Higgs field in electroweak baryogenesis.  Since
we feel that
the identification of $\dot\theta$ as an effective chemical potential is
incorrect, these models too are subject to the same criticism.  

In conclusion, we have calculated the baryon asymmetry created by a
pseudo Nambu-Goldstone boson with baryon number violating couplings in
the context of natural inflation.  We have obtained a general result for the
baryon asymmetry created by the decay of an oscillating
scalar field with baryon number
violating couplings and demonstrated explicitly that the asymmetry is
not proportional to $\theta_i$ to the first power as claimed in earlier
work.  We have also discussed our objection to the thermodynamical
generation of the baryon asymmetry using an effective chemical potential
approach in models of spontaneous baryogenesis.

{\bf Acknowledgements}
We would like to thank Fred Adams,
Dimitri Nanopoulos, Anupam Singh, and
Sridhar Srinivasan for useful discussions,
The work of A.D.~was supported in part by the Danish National Science Research
Council
through grant 11-9640-1 and in part by Danmarks Grundforskningsfond through its
support of the Theoretical Astrophysical Center.
The work of K.F.~was supported by NSF Grant No. PHY94-06745.
The work of R.R.~was supported by NSF Grant No. PHY91-16964 and by the World
Laboratory.
The work of M.S.~was supported by NSF Grant No. PHY91-16964.

{\bf Appendix A: Number Density of Produced Particles in Terms
of One Pair Production Amplitude}

Here we use the Bogolyubov transformation method to obtain
eq.~(3.2).  We show that in the lowest
order of perturbation theory,
the average number density of particle antiparticle pairs
produced by decay of the initial scalar field is given by
$$n={1\over V} \sum_{s_1,s_2} \int \widetilde{dp_1} \; \widetilde{dp_2} \;
|A|^2 \, , $$
where $A$ is the one pair production amplitude and subscripts
1 and 2 refer to the final particle and antiparticle produced.  For simplicity
we will work with scalar fields here; the generalization
to production of fermions is similar and has been performed
in ref.\r{dolgovkirilova90}.

We begin with a classical scalar field
 $\phi(t)$ coupled to a 
quantum complex scalar $\chi$:
$$ L_{int} = g \phi(t) \chi^* \chi \, . \eqno(A.1)$$
At early times $t \rightarrow - \infty$,
we take $\phi(t) = 0$ so that
$\chi$ is expanded in terms of creation and annihilation
operators,
$$\chi = \int \widetilde{dk}\;
\big[a_\bfk \exp({-i \omega t+i{\bf k \cdot x}}) + b_\bfk^\dagger 
\exp(i \omega t-i{\bf k \cdot x})\big] \, , \eqno(A.2)$$
where $\omega = \sqrt{\bfk^2 + m^2}$.  Here the commutators
are $[a_{\bfk_1}, a_{\bfk_2}^\dagger] = (2\pi)^3 2 k_1^0\delta^{(3)}
(\bfk_1-\bfk_2)$ and a similar
relation holds
for the antiparticle creation and annihilation operators $b_\bfk$.
Then, at later times, $\phi(t) \neq 0$ and eq.~(A.2) is 
replaced by
$$\chi = \int \widetilde{dk}\;
 \big[a_\bfk f_k(t) \exp(i{\bf k \cdot x}) + b_\bfk^\dagger f_k^*(t)
\exp(-i {\bf k \cdot x})\big] \, , \eqno(A.3)$$
with equation of motion
$$\big[\partial_t^2 + \bfk^2 +m^2- g \phi(t) \big]f_k(t) = 0 \, . \eqno(A.4)$$
The subscript on $f_k$, and on $\alpha_k$ and $\beta_k$ below,
refers to $|\bfk|$ and not to the momentum four vector.
For continuity at early times $f_k (t \rightarrow - \infty) = 
\exp(-i\omega t)$.
We also assume that $\phi(t) \rightarrow 0$ for 
$t \rightarrow \infty$. Then we have 
$$f_k(t \rightarrow + \infty) \rightarrow \alpha_k
e^{-i \omega t} + \beta_k e^{i \omega t} \, , \eqno(A.5)$$
so that $\chi(t)$ evolves as
$$\eqalign{
\chi(t \rightarrow + \infty) & = 
\int \widetilde{dk}\;
\big[ \exp(-i \omega t + i {\bf k \cdot x})
(\alpha_k a_\bfk + \beta_{k}^* b_{-\bfk}^\dagger ) \cr & +
\exp(i \omega t - i{\bf k \cdot x}) 
(\alpha_k^* b_\bfk^\dagger  + \beta_{k} a_{-\bfk}) \big] \, . \cr} 
\eqno(A.6)$$
One can define new creation and annihilation operators,
for particles: 
$$\tilde a_\bfk = \alpha_k a_\bfk + \beta_{k}^* b_{-\bfk}^\dagger  \, , 
\eqno(A.7a)$$
and for antiparticles: 
$$\tilde b_\bfk = \alpha_k b_\bfk + 
\beta_{k}^* a_{-\bfk}^\dagger \, . \eqno(A.7b)$$
Then the operator of final particle number is given by 
$\tilde N_\bfk = \tilde a_\bfk^\dagger \tilde a_\bfk/[2k^0 V]$.

The number of particles in the final state of momentum $\bfk$ is 
given by
$$N_\bfk = \langle 0 | \tilde N_\bfk | 0 \rangle =
|\beta_{k}|^2  
\, . \eqno(A.8)$$
Thus the total number density of produced particles is
$$n = {1 \over V} {V\over (2\pi)^3} \int d^3k \;N_\bfk = 
\int {d^3k \over (2 \pi)^3}
|\beta_k|^2 \, . \eqno(A.9)$$
This result, obtained by the method of Bogolyubov coefficients,
can be found in refs.~\r{bogolyubov,birrelldavies82}.

Now we shall calculate $\beta_k$ in perturbation theory.
Expanding $f = f_0 + f_1$, we have
$f_0 = \exp(-i \omega t)$ and the equation of motion (A.4) becomes
$$(\partial_t^2 + \bfk^2+m^2)f_1 = g \phi(t) \exp(-i \omega t) \,. 
\eqno(A.10)$$
Using the Green's function method we find
$$f_1(t) = -g \int {d \omega^\prime \over 2 \pi} {\tilde\phi (\omega^\prime-
\omega) 
\over \omega^{\prime 2} - \bfk^2-m^2} e^{-i\omega^\prime t}
\, , \eqno(A.11)$$
Taking the residue at the pole $\omega^\prime=-\sqrt{\bfk^2+m^2}=-\omega$, 
we find the coefficient of $\exp(+i \omega t)$ to be,
$$\beta_k = i g [\tilde \phi(2 \omega)]^*/2\omega \, .
\eqno(A.12))$$

Now, for comparison, let us calculate the field theory
amplitude with the interaction Lagrangian given by eq.~(A.1),
$$A = \langle k_1,\bar k_2 | i \int d^4x \;g \phi(t) \chi^* \chi
| 0 \rangle \, . \eqno(A.13)$$
Perturbatively the matrix element is easy to calculate using
eq.~(A.2), and we find
$$A = i g (2\pi)^3\delta^3(\bfk_1+\bfk_2)
\int dt\; \phi(t) \exp[i(\omega_1+\omega_2)t] \, , \eqno(A.14)$$
so that
$$|A^2| = g^2 V (2\pi)^3\delta^{(3)}(\bfk_1+\bfk_2) |\tilde \phi(\omega_1+
\omega_2)|^2 
\, . \eqno(A.15)$$
Now if we integrate over $\widetilde{dk_1}\;\widetilde{dk_2}$, 
we find that 
$$n = {1 \over V} \int  \widetilde{dk_1}\;\widetilde{dk_2}\; |A|^2
= \int {d^3k \over 
(2 \pi)^3}\; g^2{ |\tilde \phi(2\omega)|^2 \over 4 \omega^2} \, .
\eqno(A.16)$$
This is exactly eq.~(A.9) with $\beta_k$ given by eq.~(A.12).
Thus we have shown that the number density of produced
particles is given by the integral of the one pair production
amplitude squared.

{\bf Appendix B: Second Derivation of
Number Density of Produced Particles in Terms
of One Pair Production Amplitude}

Eq.~(3.2) can also be obtained using the method
presented in Sec. 4-1-1 of ref.\r{itzyksonzuber80}.
(We have ignored the higher order vacuum
graphs that give the exponential factor $\exp(-\bar n)$ in eq.~(4-23) of
ref.\r{itzyksonzuber80}.)
We have verified that we
obtain the Poisson distribution for the number of $(Q,\bar L)$ pairs and
$(\bar Q, L)$ pairs as in ref.\r{itzyksonzuber80}.
Indeed the derivation of the Poisson distribution can be done exactly along
the same lines as in ref.\r{itzyksonzuber80}. The only difference is
that in the example considered in this book the matrix element describes
the production of a single photon by an external current while in our case it
gives the amplitude for production of a {\it pair} of particles. For the
multiparticle production amplitude this gives rise to a different
normalization, namely, in the case of the production of $n$ photons the
amplitude contains the factor $1/\sqrt{n!}$ connected with identical photons
while for the case of production of $n$ pairs of $\bar Q L$ (or charge
conjugate) the amplitude contains $1/n!$. In the case of photons the
multiparticle amplitude squared contains the following $n$-dependent factors:
$|A_n^{\gamma}|^2 \sim |(n!)(1/n!)(1/\sqrt{n!})|^2 \sim 1/n!$. 
The first factor of $n!$
comes from $n!$ combinations which appear when the photon production operator
act on the multiphoton state $\langle k_1,k_2.., k_n | | (a^+_k)^n$. 
The factor of $1/n!$
comes from the expansion of the action $S= \exp(i\int d^4x A^\mu J_\mu)$,
and the factor of $1/\sqrt{n!}$ comes from the normalization of the $n$-photon
state. So the net result is proportional to $1/n!$, which is exactly what is
needed to get the Poisson distribution $p_n = \exp(-\bar n) \bar n^n /n!$.
In the case of the production of $n$-pairs, we have the same $1/n!$ from the
expansion of the action, but now we get $1/n!$ coming from the normalization
and not $1/\sqrt{n!}$ as before.  However, the action of the product of the
creation operators of $Q$ and $\bar L$, which can be symbolically written as
$(a^+_Q b^+_L)^n$, gives now an overall factor of $n!$ from the action of,
say, $(a_Q^+)^n$, as above, and also the sum
of $n!$ equal but not interfering terms, each of them being proportional to a
different delta-function of the momenta, $\delta (p_{Q_j}+p_{L_k})$. Thus in
the matrix  element squared we will get the same overall factor $1/n!$
which is necessary for the Poisson distribution.

{\bf Appendix C: Calculation of Baryon Asymmetry}

Here we calculate the lowest order nonzero
contribution to the baryon asymmetry; we derive eq.~(3.13)
from eqs.~(3.9) and (3.10). As our starting point, we have
$$n_b - n_{\bar b} =  {g^2 f^2 \over \pi^2}
\int d\omega \, \omega^2 \Biggl[ {\tilde \theta(2 \omega) \,
\left[\widetilde{\theta^2}(2 \omega)\right]^*\over 2i} 
\, + {\rm h.c.}\Biggr], \eqno(C.1)$$
where 
$$\tilde \theta (2 \omega) =  \int_{-\infty}^{\infty} dt \, e^{2 i \omega t} \,
\theta (t) \, \eqno(C.2)$$
and
$$\widetilde{\theta^2}(2\omega) =  \int_{-\infty}^{\infty} dt \, e^{2i\omega t} 
\, \theta^2(t) \, . \eqno(C.3)$$  
Using eq.~(3.8), we find that
$$\tilde \theta (2 \omega ) = {\theta_i \over 4 i \omega}
\bigl[ {(- \Gamma/2 +im_R) \over (- \Gamma/2 + i m_R 
+ 2 i \omega)} - {(\Gamma/2 +im_R) \over (- \Gamma/2 -im_R +2i \omega)}\bigr]
\,  \eqno(C.4)$$
and
$$\widetilde{\theta^2}(2 \omega)^* = - {\theta_i^2 \over 4 i \omega}
\bigl[ {(im_R + \Gamma/2) \over 2im_R +2i \omega + \Gamma}
+ {( -im_R + \Gamma/2) \over 2i \omega -2im_R + \Gamma}
+ { \Gamma \over 2i \omega + \Gamma} \bigr] \eqno(C.5)$$
Thus
$$\eqalign{
\tilde \theta \widetilde{\theta^2}^* =
&{ \theta_i^3 \over 16 \omega^2} \Biggl[
{(-m_R^2 - \Gamma^2/4) \over (2im_R + 2i \omega + \Gamma)
(2 i\omega + i m_R - \Gamma/2) }
+ { (m_R^2 - \Gamma^2/4 + \Gamma i m_R )
\over (2 i \omega - 2im_R + \Gamma) (2 i \omega + im_R - \Gamma/2)}\cr
& + {\Gamma (im_R - \Gamma/2) 
\over (2 i \omega + \Gamma) (2 i \omega + i m_R - \Gamma/2)} 
- {-m_R^2  + i \Gamma m_R +\Gamma^2/4
\over (2im_R +2i\omega + \Gamma) (2 i \omega -im_R - \Gamma/2)}\cr
&- { m_R^2 + \Gamma^2/4
\over (2 i \omega - 2i m_R + \Gamma) (2 i \omega - im_R - \Gamma/2)}
- { \Gamma (im_R + \Gamma/2)
\over (2 i \omega + \Gamma) (2i \omega - im_R - \Gamma/2)}
\Biggr] \, . \cr} \eqno(C.6)$$
Now we must integrate each of the terms in eq.~(C.6)
as indicated in eq.~(C.1).  The lower limit of the integral is
$m_Q+m_L\ll m_R$ and we use $\Gamma\ll m_Q + m_L$.
We find that the first term cancels with its hermitian conjugate, the third
and sixth terms 
are 0, the second and fourth terms 
cancel each other,
and the fifth term plus its hermitian conjugate
is responsible for the final result given in eq.~(3.13),
$$\eqalign{
n_B\equiv n_b-n_{\bar b}&={g^2\over 16\pi} m_Rf^2\th_i^3 \cr
&={1\over 2}\Gamma f^2 \th_i^3\cr}\, .\eqno(C.7)$$

\medskip

{\bf Appendix D: The Effects of Mixing in the $Q,L$ Basis}

We will consider the decay of $\theta$ to a $Q \bar L$ pair
(superscript $1$ for this decay channel), and the
decay of $\theta$ to a $\bar Q L$ pair (superscript $2$
for this decay channel).
For the first decay channel,
from eq.~(3.16) we see that a $Q$ produced at the time $t=0$
is given by
$$\psi(0)=Q = s \psi_1 + c \psi_2 \, , \eqno(D.1a)$$
where $$c = {1 \over \sqrt{1 + \epsilon^2}} \quad
{\rm and} \quad s = { \epsilon \over \sqrt{1 + \epsilon^2}} \, .
\eqno(D.1b)$$
Similarly,
$$\bar \chi(0) =\bar L= c \bar \psi_1 -s \bar \psi_2 \, . \eqno(D.2)$$
We will let the fields $\psi$ and $\chi$
evolve in time, mixing their $Q$ and $L$ components
as they travel.
The time evolution of $\psi(t)$ can be modeled as follows:
$$\psi(t) = (s e^{-i \Delta \omega t} \psi_1 + c \psi_2 ) 
\exp(-i\omega_2 t)
\, , \eqno(D.3)$$
where $\Delta \omega = \omega_1 - \omega_2$.
We now wish to ask the question: what is the $Q$ content
at some time $t$ of the field $\psi$ which was initially pure $Q$?
Using eq.~(3.16), we can write eq.~(D.3) as
$$\psi(t) = \bigl[(c^2 + s^2 e^{-i \Delta \omega t}) Q
- s c (1 - e^{-i \Delta \omega t}) L\bigr] \exp(-i \omega_2 t)
\, . \eqno(D.4)$$
The quark content is given by the magnitude squared of the coefficient
of the first term, so that
$$n^{(1)}_Q(t) = \bigl[c^4 + s^4 + 2 c^2 s^2 {\rm cos} \Delta \omega t 
\bigr] {1 \over V} \sum_{s_Q,s_{\bar L}} 
\int \widetilde{dk_Q}\; \widetilde{dk_{\bar L}}\;
|A_{Q\bar L}|^2 \, . \eqno(D.5)$$
Similarly, from the same decay process $\theta \rightarrow Q + \bar L$, 
the $\bar L$ that is produced can convert to a $\bar Q$ so that
we have
$$n^{(1)}_{\bar Q}(t) = {1\over V} \sum_{s_Q,s_{\bar L}}  
2 s^2 c^2 (1 - {\rm cos} \Delta \omega t) \int \widetilde{dk_Q}\;
\widetilde{dk_{\bar L}} \; |A_{Q \bar L}|^2
\, . \eqno(D.6)$$
{}From $\theta \rightarrow L \bar Q$, one can obtain $Q$ at
a later time from oscillations of either the $L$ or the
$\bar Q$ and find contributions:
$$n^{(2)}_{\bar Q}(t) = \bigl[c^4 + s^4 + 2 c^2 s^2 {\rm cos} \Delta \omega t 
\bigr] {1 \over V} \sum_{s_L,s_{\bar Q}}  
\int \widetilde{dk_L}\;\widetilde{dk_{\bar Q}}\; 
|A_{L \bar Q}|^2 \,  \eqno(D.7)$$
and
$$n^{(2)}_Q(t) = {1\over V} \sum_{s_L,s_{\bar Q}}
2 s^2 c^2 (1 - {\rm cos} \Delta \omega t) \int
\widetilde{dk_L}\; \widetilde{dk_{\bar Q}} \; |A_{L \bar Q}|^2
\, . \eqno(D.8)$$
Thus the baryon asymmetry at any time $t$ is
$$\eqalign{
n_B(t) &= n^{(1)}_Q(t) + n^{(2)}_Q(t) 
- n^{(1)}_{\bar Q}(t) - n^{(2)}_{\bar Q}(t) \cr
&= \big[(c^2 - s^2)^2 + 4 s^2 c^2 {\rm cos} \Delta \omega t \bigr]
\sum_{s_L,s_{ Q}}\bigl( |A_{Q\bar L }|^2 - |A_{L\bar Q }|^2  \bigr) \cr
&= \Biggl[ \Biggl({1 - \epsilon^2 \over 1+ \epsilon^2}\Biggr)^2
+ 4 s^2 c^2 {\rm cos} \Delta \omega t \Biggr]
 \sum_{s_L,s_{Q}}\bigl( |A_{Q\bar L }|^2 - |A_{L\bar Q }|^2 \bigr) \, . \cr} 
\eqno(D.9)$$
One can see that the baryon asymmetry oscillates in time as
a cosine about the average value.
When one takes a time average, the cosine term averages to 
zero, and one reproduces the result in eq.~(3.28),
$$n_B={1\over 2}\Gamma f^2 \th_i^3
\Biggl({1-\epsilon^2\over 1+\epsilon^2}\Biggr)^2
\,. \eqno(D.10)
$$

Our derivation above assumes in eqs.~(D.5-D.8) that all $Q\bar L$ pairs
and all $L\bar Q$ pairs were
produced at the same time.  If one considers that all pairs are not
produced at the same time then an average over all pairs would also
cancel the ${\rm cos}\Delta \omega t$ term in eqs.~(D.5-D.8).

\references

\hyphenation{Amsterdam}

\refis{cohkap8788}A. G. Cohen and D. B. Kaplan, \pl B199, 1987, 251;
\np B308, 1988, 913.

\refis{freeseetal90}K. Freese, J. A. Frieman and A. V. Olinto,
\prl 65, 1990, 3233.  

\refis{steinhardtturner84}P. J. Steinhardt and M. S. Turner,
\prd 29, 1984, 2162.

\refis{adamsfreeseguth91}F. C. Adams, K. Freese and A. H. Guth,
\prd 43, 1991, 965.

\refis{models}Natural inflation can be realised 
in many realistic particle physics
models in which a Nambu-Goldstone boson acquires a potential of the form
in eq.~(1.1).  In a class of $Z_2$ symmetric models
the combination of terms like $m_1 \bar \psi \psi$
and $m_0 (\bar \psi \psi e^{i\theta} + h.c.)$ can give rise to a
potential as in eq.~(1.1) for $\theta$
with $\L^2\sim m_0 m_1$.  These ``schizon'' models are further described
in ref.\r{changetal85,hillross88,friemanetal92}.
In superstring models, non-perturbative effects in the hidden sector can
give rise to fermion condensates and, consequently, a potential for the
model independent axion (the imaginary part of the dilaton field).  In
refs.\r{deribnil,dinerosewit}, the hidden $E_8^\prime$ sector in
$E_8\times E_8^\prime$ heterotic string theory becomes strongly
interacting, generating gaugino condensates that lead to SUSY breaking
and a potential for the model independent axion.  Problems with the
stability of the dilaton potential in such models has prompted others to
consider multiple gaugino condensation models which give a suitable
potential for the axion\r{krasnikov87,caslalmunross90,KDLP}.
Fermion condensates in
technicolor theories can also give rise to potentials of the form above
for fields coupled to the fermions.
Also, a theory
with an antisymmetric tensor field $B_{\mu\nu}$ (which arises,
for example, in string theory) with a field strength
$$H^{\mu\nu\lambda}=\partial^\mu B^{\nu\lambda}+\partial^\lambda
B^{\mu\nu} +\partial^\nu B^{\lambda\mu}$$
has an effective action which can be expressed in terms of a
scalar field with a potential of the form in eq.~(1.1)\r{ovrutthomas9192}.
In a variant of these models, the tensor field can be coupled to a 
fundamental real scalar field $u$ with the symmetry breaking potential
of the form
$V(u)=({\lambda^\prime\over 4!})(u^2-6m^2/\lambda^\prime)^2$.  
This also leads to a potential as in eq.~(1.1) for the scalar $\theta$.

\refis{adamsetal}F. C. Adams, J. R. Bond, K. Freese, J. A. Frieman and
A. V. Olinto, \prd 47, 1993, 426.  

\refis{krasnikov87}N. V. Krasnikov, \pl B193, 1987, 37.

\refis{caslalmunross90}J. A. Casas, Z. Lalak, C. Munoz and G. G. Ross,
\np B347, 1990, 243.

\refis{KDLP}V. Kaplunovsky, L. Dixon, J. Louis and M. Peskin
(unpublished); L. Dixon, in {\it The Rice Meeting}, Proceedings of the
Annual Meeting of the Division of Particles and Fields of the APS,
Houston, Texas, 1990, edited by B. Bonner and H. Miettinen (World
Scientific, Singapore, 1990); J. Louis, in {\it The Vancouver
Meeting-Particles and Fields '91}, Proceedings of the Joint Meeting of
the Division of Particles and Fields fo the American Physical Society
and the Particle Physics Division of the Canadian Association of
Physicists, Vancouver, 1991, edited by D. Axen, D. Bryman and M. Comyn
(World Scientific, Singapore, 1992).

\refis{dolgovfreese}A. D. Dolgov and K. Freese, \prd 51, 1995, 2693.

\refis{bogolyubov}N. N. Bogolyubov, {\it Lectures on Quantum Mechanics},
(Naukova Dumka, Kiev, 1970).  See also A. A. Grib, S. G. Mamayev, and 
V. M. Mostepanenko, {\it Vacuum Effects in Strong Fields}, (Friedman Lab.
Publishing, St. Petersburg, 1994).

\refis{birrelldavies82}N. D. Birrell and P. C. W. Davies, {\it Quantum
Fields in Curved Space}, (Cambridge University Press, Cambridge, 1982).

\refis{itzyksonzuber80}C. Itzykson and J. Zuber, {\it Quantum Field Theory},
(McGraw-Hill Book Company, 1980).

\refis{changetal85}D. Chang, R. N. Mohapatra and S. Nussinov,
\prl 55, 1985, 2835.

\refis{hillross88}C. T. Hill and G. G. Ross, \pl B203, 1988, 125;
\np B311, 1988, 253.

\refis{friemanetal92}J. A. Frieman, C. T. Hill and R. Watkins, \prd 46,
1992, 1226.

\refis{deribnil}J. P. Derendinger, L. E. Ibanez and H. P. Nilles, \pl
B155, 1985, 65.

\refis{dinerosewit}M. Dine, R. Rohm, N. Seiberg and E. Witten, \pl
B156, 1985, 55.

\refis{ovrutthomas9192}B. Ovrut and S. Thomas, \pl B267, 1991, 227;
\pl B277, 1992, 53.

\refis{dolgovkirilova90}A. D. Dolgov and D. P. Kirilova, Sov. J. Nucl.
Phys. {\bf 51} (1990) 172.

\refis{ewspbar}M. Dine, P. Huet, R. Singleton Jr. and L. Susskind, 
{\it Phys. Lett.} {\bf B257} (1991) 351; 
A. G. Cohen, D. B. Kaplan and A. E. Nelson, 
{\it Phys. Lett.} {\bf B263} (1991) 86; 
M. Dine, P. Huet and R. Singleton Jr., 
{\it Nucl. Phys.} {\bf B375} (1992) 625;
M. Dine and S. Thomas,
{\it Phys. Lett.} {\bf B328} (1994) 73; 
T. Prokopec, R. Brandenberger, A. C. Davis and M. Trodden,
{\it hep-ph/9511349} (1995),
T. Prokopec, R. Brandenberger and A.-C. Davis,
{\it hep-ph/9601327} (1996).

\refis{stecker89others}F. W. Stecker, {\it Nucl.Phys. (Proc. Suppl.)} 
{\bf B10} (1989) 93.
For a more recent discussion see, e.g., A.D. Dolgov, {\it Hyperfine 
Interactions} {\bf 76} (1993) 93 and A. De Rujula, {\it Nucl. Phys. (Proc. 
Suppl.)} {\bf B48} (1996) 514.

\refis{voloshin}M. Voloshin, private communication.

\endreferences

\end